\documentclass[aps,preprintnumbers,amsmath,amssymb,amsfonts, preprint,tightenlines]{revtex4}

\usepackage{bm}

\newcommand{\dz}{\partial_z}
\newcommand{\dzb}{\partial_{\bar{z}}}
\newcommand{\psip}{\psi_+}
\newcommand{\psim}{\psi_-}
\newcommand{\betap}{\beta_+}
\newcommand{\betam}{\beta_-}
\newcommand{\ps}[3]{\psi_{#1}^{#2}({#3})}
\newcommand{\be}[3]{\beta_{#1}^{#2}({#3})}
\newcommand{\syp}{\mathfrak{sp}}
\newcommand{\su}{\mathfrak{su}}
\newcommand{\so}{\mathfrak{so}}
\newcommand{\gl}{\mathfrak{gl}}
\newcommand{\osp}{\mathfrak{osp}}
\newcommand{\uone}{\mathfrak{u}}
\newcommand{\g}{\mathfrak{g}}

\def\Jbar{\bar{J}}

\def\bar{\overline}
\def\hat{\widehat}
\def\*{\star}
\def\[{\left[}
\def\]{\right]}
\def\({\left(}		
\def\){\right)}

\def\2pi{\hbox{$2\pi i$}}

\def\dsl{\raise.15ex\hbox{/}\kern-.57em\partial}
\def\Dsl{\,\raise.15ex\hbox{/}\mkern-.13.5mu D}
\def\2pi{\hbox{$2\pi i$}}

\def\dsl{\raise.15ex\hbox{/}\kern-.57em\partial}
\def\Dsl{\,\raise.15ex\hbox{/}\mkern-.13.5mu D}
\font\numbers=cmss12
\font\upright=cmu10 scaled\magstep1
\def\stroke{\vrule height8pt width0.4pt depth-0.1pt}
\def\topfleck{\vrule height8pt width0.5pt depth-5.9pt}
\def\botfleck{\vrule height2pt width0.5pt depth0.1pt}
\def\Zmath{\vcenter{\hbox{\numbers\rlap{\rlap{Z}\kern
0.8pt\topfleck}\kern 2.2pt
                   \rlap Z\kern 6pt\botfleck\kern 1pt}}}
\def\Qmath{\vcenter{\hbox{\upright\rlap{\rlap{Q}\kern
                   3.8pt\stroke}\phantom{Q}}}}
\def\Nmath{\vcenter{\hbox{\upright\rlap{I}\kern 1.7pt N}}}
\def\Cmath{\vcenter{\hbox{\upright\rlap{\rlap{C}\kern
                   3.8pt\stroke}\phantom{C}}}}
\def\Rmath{\vcenter{\hbox{\upright\rlap{I}\kern 1.7pt R}}}
\def\Z{\ifmmode\Zmath\else$\Zmath$\fi}
\def\Q{\ifmmode\Qmath\else$\Qmath$\fi}
\def\N{\ifmmode\Nmath\else$\Nmath$\fi}
\def\C{\ifmmode\Cmath\else$\Cmath$\fi}
\def\R{\ifmmode\Rmath\else$\Rmath$\fi}

\def\barray{\begin{eqnarray}}
\def\earray{\end{eqnarray}}
\def\beq{\begin{equation}}
\def\eeq{\end{equation}}
\begin{document}

    \bibliographystyle{apsrev}

    \title {Super Spin-Charge Separation for class {\bf A}, {\bf C}, and {\bf D}
disorder}
                
    \author{Andr\'e LeClair}
                \email[email:]{arl4@cornell.edu}
    \author{Dean J. Robinson}
                \email[email:]{djr233@cornell.edu}
    \affiliation{Department of Physics, Cornell University, Ithaca, N.Y.}
    \date{July 2008}
   
\bigskip\bigskip\bigskip
 
    \begin{abstract}
        
We prove  versions of super spin-charge separation 
for all three of the symmetry groups SU(N), Sp(2N), and SO(N)
of  disordered Dirac fermions in $2+1$
dimensions,  which  involve  the supercurrent-algebras 
$\gl (1|1)_{N}$, $\osp(2|2)_{-2N}$, and $\osp(2|2)_N$ respectively.   
For certain restricted classes of disordered potentials, the latter
supercurrent algebra based conformal field theories can arise
as non-trivial low energy fixed points.   For all cases with such
a fixed point,  we compute
the density of states exponents as a function of $N$.   

\end{abstract}
\maketitle

\section{Introduction}

The Wigner-Dyson classification of random hamiltonians based on 
time-reversal symmetry consists of three universality classes of ensembles:
the GUE, GSE, and GOE with U(N), Sp(2N) and O(N) symmetry respectively.  
Extending this classification to incorporate particle-hole and chiral symmetries,
Altland-Zirnbauer found 10 classes \cite{Altland:1997pr,Zirnbauer:1996mp}.  
This  kind of classification was studied  specifically for  disordered Dirac
fermions in $2d$  where a minor refinement consisting of
13 classes was found \cite{Bernard:2002ar}.
All of these universality classes are characterized by 
the symmetries U(N), Sp(2N), or O(N), or two copies of these, 
and following Refs. \cite{Altland:1997pr,Zirnbauer:1996mp}, we will refer to them generally
as class {\bf A, C,} or {\bf D} respectively.

Disordered Dirac fermions are especially interesting as possible models
of Anderson transitions in $2+1$ dimensions.  (For a recent review, see Ref. \cite{Mirlin:2007ar}). 
Using  the supersymmetric method for performing disorder averages,
the disorder averaged effective action consists of a $2d$  field theory
of Dirac fermions and ghosts with marginal current-current perturbations, where the
coupling constants measure the strengths of the various disordered 
potentials \cite{Bernard:1995cc,Mudry:1996qy}.  One is then primarily  interested in
identifying the possible quantum critical points as fixed points of
the renormalization group (RG).   This has turned out to be a rather challenging
problem.   The main source of difficulty is due to the marginality 
of the perturbations,  which generally do not reveal  fixed points
in perturbation theory.   Another problem is  that models typically have several couplings;  
for example the class {\bf A} model corresponding to the network
model for the quantum Hall transition  has three independent couplings,
and the higher loop corrections to the beta function do not reveal
a fixed point \cite{Bernard:2002kx}.     The generic 1-loop beta functions
for all classes of disordered Dirac fermions  were computed in Refs. \cite{Bocquet:2000dq,Serban:2002df}.

For ordinary marginal current-current perturbations in the non-disordered
context,  a non-perturbative mechanism for obtaining
a low energy fixed point exists based on spin-charge separation.  
The prototype for this mechanism is the fixed point for the $1d$ Hubbard 
model \cite{Affleck:1988lh};   in fact the low energy fixed point of the $1d$ Hubbard
model is very difficult to understand without  using the spin-charge separation.
 Extending the discussion to the $N$-component version,
where $N=2$ is the Hubbard model,  the stress-tensor of $N$ free Dirac fermions
can be decomposed as: 
\begin{equation}
	\label{eqn:HM}
	T_{\rm free} =  T_{\uone(1)}  +  T_{\su(N)_1}
\end{equation}
where $T_{\su(N)_k}$ is the Sugawara stress tensor based on the $\su(N)$ current
algebra at level $k$ and $T_{\uone(1)}$ is the stress tensor for a single scalar field
at a particular radius (see e.g. Ref. \cite{Ginsparg:1988ui}).    
For $N=2$ the above decomposition is a precise statement of the spin-charge
separation in the $1d$ Hubbard model.    More generally,  given 
a  decomposition of the above form with $\uone (1)$ and $\su(N)$ replaced by
the two current algebras $\g$ and $\g'$,  consider the action
\begin{equation}
\label{eqn:AP}
\mathcal{S} = \mathcal{S}_{\rm free}  +  \int  \frac{d^2 z}{4\pi}  \(  g  J \cdot 
 \Jbar + g'  J'  \cdot \Jbar' \)
\end{equation} 
where  $J$ and $\Jbar$  are respectively the left and right  currents for $\g$
with $J\cdot \Jbar$ invariant, and similarly for $J' , \Jbar'$. 
If one of the couplings,  say $g$,  is marginally relevant,  then one can
argue that the $\g$ sector is massive and gapped out in the RG  flow to low energies. 
If on the other hand $g'$ is marginally irrelevant,  the sector 
$\g'$ survives the flow  and the low energy fixed point is the conformal field theory
$\g'$.   In the application of this idea to the Hubbard model, 
the charge U(1) degrees of freedom are actually part of
an additional SU(2) and correspond to the massive sector that is gapped out in
the flow.  The spin-SU(2) perturbations are marginally irrelevant 
and the low energy  fixed point is the current algebra $\su(2)_1$
(see Ref. \cite{Affleck:1988lh} for details).

In the disordered systems case, although it  is certainly possible that some
as yet unknown non-perturbative mechanism is responsible for expected critical  points,
since the above mechanism is the only currently 
known  manner to  obtain non-trivial  fixed points
from current-current interactions,  it is worthwhile to explore its analog
in the disordered Dirac fermion context.  
In this article, we show that there exists a super spin-charge separation  for all 
of the symmetry groups which arise in the classification of disordered
Dirac  Hamiltonians.   This is achieved by finding explicitly the decompositions
for each case. For  Sp(2N), the  decomposition 
\begin{equation}
	\label{eqn:SPS}
	T_{\textrm{free}} = T_{\osp(2|2)_{-2N}} + T_{\syp(2N)_0}
\end{equation}
 was proven in Ref. \cite{Bernard:2001np}.   The special case of SU(2)=Sp(2) was 
first considered 
in Refs. \cite{Bernard:2000vc,Bhaseen:2001ty}. In the following we prove the 
previously postulated \cite{LeClair:2007aj} SU(N) decomposition:
\beq
\label{sun}
 T_{\textrm{free}} = T_{\gl(1|1)_{N}}+ T_{\su(N)_0}
\eeq
We further present and prove a similar, new decomposition  for the SO(N) case: 
\beq
\label{son}
T_{\textrm{free}} = T_{\osp(2|2)_N} + T_{\so(N)_0}
\eeq
In all the above cases $T_{\rm free}$ represents the free conformal field theory
of $N$ Dirac fermions and ghosts. 

The above arguments show how the supercurrent algebra 
theories $\osp(2|2)_{-2N}, \osp(2|2)_N$ 
and $\gl(1|1)_N$ can arise as non-trivial fixed points of $N$-component disordered Dirac
fermions, as least  when  the effective action is of the form (\ref{eqn:AP}) with
$\g$ the bosonic level-0 current algebra $\su(N)_0$, $\syp(2N)_0$ or $\so(N)_0$,
 and $\g'$ is the supercurrent algebra  $\gl(1|1)_N, \osp(2|2)_{-2N}$ or $\osp(2|2)_N$ respectively, since, for positive couplings $g'$,
the current-current perturbation for $\su(N), \syp(2N)$  or $\so(N)$ is marginally
relevant,  whereas the coupling for the supergroup $\osp(2|2)$ or $\gl(1|1)$ is
irrelevant.  It should be kept in mind that 
 the generic model in each universality class has additional
forms of disorder beyond that corresponding to these two couplings,  and 
consequently different fixed points from these are in principle 
 possible when the additional kinds
of disorder are taken into account.   This possibility
will  not be studied in this paper,  however we point out that 
additional forms of disorder perhaps can be incorporated as 
additional perturbations of the $\g'$ theory, as suggested in
Ref. \cite{LeClair:2007it}.

Since the Sugawara stress tensors in the above super spin-charge separations
represent stress tensors for interacting theories,  the low energy fixed
points have non-trivial exponents.   The super spin-charge separation
implies fields in the free theory can be factored into fields in
the two decoupled sectors.     Letting $\Delta$ denote the 
left-moving conformal scaling dimension, then
\beq
	\label{eqn:CSD}
	\Delta_{\rm free} =  \Delta_{\g}  + \Delta_{\g'} 
\eeq
If $\g$ is gapped out in the flow to low energies,  then
the scaling dimension at the fixed point is given by
$\Delta_{\g'}$. 
For instance,  the fundamental Dirac fields have $\Delta_{\rm free} = 1/2$,
which become non-trivial functions of $N$ at the low energy fixed point.      
As an example of the utility of these super-spin 
charge separations, the density of states exponents for arbitrary N are calculated.

In the next section we present detailed proofs of the super spin-charge
decompositions for SU(N) and SO(N).   In section III we show how the super-spin
charge separations  are preserved under certain restricted classes of disorder.   In section \ref{sec:DSE} the density
of states exponents are presented.

\section{Proofs of the decompositions}
\label{sec:PD}

Consider $N$ free, massless, Dirac fermions in $2$  euclidean dimensions, each 
with a global U(1) charge. As usual, in the euclidean light-cone 
coordinates $z = (x +iy)/\sqrt{2}$, $\bar{z} = z^*$, the Dirac  fermions may be written as 
$\psi^i_\pm = \psi^i_{\pm}(z)$ and $\bar{\psi}^i_{\pm} = \bar{\psi}^i_\pm(\bar{z})$, where the subscript indicates the sign of the $U(1)$ charge,
and $i=1,..,N$. 
 (Henceforth, for the sake of brevity, the anti-holomorphic or right-moving 
fermions and their corresponding currents and terms are omitted.) 
Together with their bosonic ghost partners $\beta_\pm$, these fermions form a 
two-dimensional free superconformal field theory with action
\begin{equation}
	\label{eqn:AFCFT}
	\mathcal{S}_{\textrm{free}} = \int \frac{d^2z}{4\pi}\Big(\psim^i\dzb\psip^i + \betam^i\dzb\betap^i\Big)
\end{equation}
 Summation over repeated indices will always be implied.  The action 
has the maximal symmetry corresponding to the supergroup OSp(2N$|$2N).

The stress-energy tensor for the free theory is
\begin{equation}
	\label{eqn:SETCFT}
	T_{\textrm{free}} = -\frac{1}{2}\colon\!\!\psim^i\dz\psip^i + \psip^i\dz\psim^i + \betam^i\dz\betap^i - \betap^i\dz\betam^i\colon
\end{equation}
where `$\colon\!\!\ldots \colon$' denotes normal-ordering. Henceforth we drop this notation, and instead normal-ordering is implied in the usual contexts. The propagators for the fermions and ghosts are
\begin{align}
\langle\ps{-}{i}{z}\ps{+}{j}{w}\rangle & = \langle\ps{+}{i}{z}\ps{-}{j}{w}\rangle = \delta^{ij}/(z-w)~,\notag\\
\langle\be{+}{i}{z}\be{-}{j}{w}\rangle & = -\langle\be{-}{i}{z}\be{+}{j}{w}\rangle =  \delta^{ij}/(z-w)~. \label{eqn:FGP}
\end{align}
As expected, it is easy to check that $T_\textrm{free}$ has central charge $c=0$, 
and the conformal dimension  of the primary fields $\psi_\pm$, 
$\beta_\pm$ is $\Delta_{\textrm{free}} = 1/2$.

In order to derive the SU(N) and SO(N) decompositions 
 claimed above, we require a representation of the affine current algebras $\su(N)$ and $\so(N)$ both at level $0$. A convenient choice is as follows. Consider a (semisimple) Lie group $G$ with an $N$-dimensional irreducible representation under which $\psi_\pm^i$ and $\beta_\pm^i$ may simultaneously transform as $N$-tuples. Let the generators of this representation be $t^a$, $a = 1,\ldots, |G|$ and consider the currents
\begin{equation}
	J^a \equiv \psim^i t^a_{ij} \psip^j~.
\end{equation}
By Eqs. (\ref{eqn:FGP}) and the anticommutativity of $\psi_\pm$, these currents have operator product expansion (OPE)
\begin{equation}
	\label{eqn:DKMA}
	J^a(z)J^b(0) \sim \frac{\delta^{ab}C(N)}{z^2} + \frac{if^{abc}}{z} J^c(0) + \ldots~,
\end{equation}
where $C(N)$ is the Casimir for the representation $t^a$ and $f^{abc}$ are the structure constants of $G$. By definition, the currents $J^a$ thus form a representation of the affine (Kac-Moody) current algebra of $G$ at level $2C(N)/\alpha^2$, where $\alpha^2$ is the square length of the longest root of the representation $t^a$ (see e.g. Ref. \cite{Ginsparg:1988ui}). In this paper, the current algebra of $G$ at level $k$ is denoted $\mathfrak{g}_k$ and we choose the 
normalization $\alpha^2 =2$,  
so that $J^a$ form a representation of $\mathfrak{g}_{C(N)}$. Similarly, the ghost currents $\betam^i t^a_{ij} \betap^j$ form a representation of $\mathfrak{g}_{-C(N)}$, and it follows that
\begin{equation}
	\label{eqn:GLAC}
	L^a \equiv \psim^it^a_{ij}\psip^j + \betam^it^a_{ij}\betap^j,
\end{equation}
is a representation of $\mathfrak{g}_0$.

Let us now find explicit expressions for the $\su(N)_0$ and $\so(N)_0$ stress-energy tensors, in terms of the fermion and ghost fields. Employing the Sugawara construction and the currents $L^a$, the stress-energy tensor for current algebra $\mathfrak{g}_0$ is
\begin{align}
	T_{\mathfrak{g}_0}(z) 
	& = \frac{1}{C_2(G)}\sum_{a=1}^{|G|}\colon \!\! L^a(z)L^a(z) \colon \notag\\
	& = \frac{1}{C_2(G)}\lim_{z \to w}\sum_{a=1}^{|G|}L^a(z)L^a(w) \label{eqn:GSSET}
\end{align}
where $C_2(G)$ is the quadratic Casimir of the adjoint representation of $G$ and noting from Eq. (\ref{eqn:DKMA}) that $\colon\!\! L^aL^a \colon = L^aL^a$, since the level of the algebra is zero here. 

Applying the prescription (\ref{eqn:GSSET}), it is clear from Eq. (\ref{eqn:GLAC}) that the direct product $t^a_{ij}t^a_{kl}$ will arise in the expression for $T_{\mathfrak{g}_0}$. The stress-energy tensor should be independent of the choice of the matrices $t^a$, and one may find identities for these products in terms of Kronecker deltas. In the $\su(N)$ case, the fermion and ghost fields clearly transform under the fundamental representation, the generators of which are the $N^2 -1$ traceless Hermitian $N\times N$ matrices. Together with the definition of the quadratic Casimir, these properties of the generators produce the result
\begin{equation}
	\label{eqn:SUGI}
	t^a_{ij}t^a_{kl}  \equiv C_2(N)\frac{N}{N^2-1}\bigg(\delta_{il}\delta_{jk} - \frac{1}{N}\delta_{ij}\delta_{kl}\bigg)~,
\end{equation}	
where $C_2(N)$ is the quadratic Casimir of the fundamental representation. Under the normalization $\alpha^2 =2$  we have
\begin{equation}
	\label{eqn:SUCI}
	C_2(G) = 2N,~C(N) =1,~ \mbox{and}~C_2(N) = (N^2-1)/N
\end{equation}
for the $\su(N)$ algebra.

In the $\so(N)$ case, the fermion and ghost fields transform under the vector representation, the generators of which are the $N(N-1)/2$ real antisymmetric $N \times N$ matrices. One may show that these generators satisfy
\begin{equation}
	\label{eqn:SOGI}
	t^a_{ij}t^a_{kl} \equiv C_2(N)\frac{1}{N-1}\Big(\delta_{il}\delta_{jk} - \delta_{ik}\delta_{jl}\Big)~,
\end{equation}
and  we have
\begin{equation}
	\label{eqn:SOCI}
	C_2(G) = 2(N-2),~C(N) = 2,~\mbox{and}~C_2(N) = N - 1~.
\end{equation}

Applying the Wick theorem to the operator products in Eq. (\ref{eqn:GSSET}) using Eqs. (\ref{eqn:FGP}), it follows from Eqs. (\ref{eqn:SUGI}), and (\ref{eqn:SUCI}) that the $\su(N) _0$ stress energy tensor is
\begin{align}
 T_{\su(N)_0}	& = \frac{1 - N^2}{2N^2}\bigg(\psim^i\dz\psip^i + \psip^i\dz\psim^i + \betam^i\dz\betap^i - \betap^i\dz\betam^i\bigg) \notag \\
	& ~~~~+ \frac{1}{2N^2}\bigg((N-1)(\betap^i\betam^i)^2  -
 (N+1)(\psip^i\psim^i)^2  \notag \\
	 & ~~~~~~~~~~ - 2(\psim^i\psip^i)(\betam^j\betap^j) + 2N(\betap^i\psim^i)(\betam^j\psip^j)\bigg). \label{eqn:SETSU}
\end{align}
The derivative terms arise from the Taylor expansion of $\psi_{\pm}(z)$ about $w$, before taking the limit $z \to w$. Similarly, from Eqs. (\ref{eqn:SOGI}) and (\ref{eqn:SOCI}) one finds 
\begin{align}
	 T_{\so(N)_0} 
	& = \frac{1-N}{2(N-2)}\bigg(\!\psim^i \dz\psip^i+ \psip^i \dz\psim^i + \betam^i\dz\betap^i - \betap^i\dz\betam^i\!\bigg) \notag\\
	&~~~ + \frac{1}{2(N-2)}\bigg( (\betap^i\betam^i)^2 - (\betap^i\betap^i)(\betam^j\betam^j) - (\psip^i\psim^i)^2   \notag \\  
	&~~~~~~~~~~
 + 2(\psim^i\betap^i)(\psip^j\betam^j) - 2(\psim^i\betam^i)(\psip^j\betap^j)\bigg)~. \label{eqn:SETSO}
\end{align}
Restoring the normal-ordering notation, note that there should also be a quartic term of the form $\colon\!\!(\psip^i\psip^i)(\psim^j\psim^j)\colon$ in Eq. (\ref{eqn:SETSO}), arising from the $J^aJ^a$ product. However, this term is omitted since the anticommutativity of the fermion fields guarantees that $\colon\!(\psip^2)(\psim^2)\colon\! = \colon\!\psip^2\colon\!\colon\!\psim^2\colon\! = 0$.

It has been deduced previously via counting the field degrees of freedom, together with consideration of the scaling dimensions, that $T_{\textrm{free}}$ factors into stress energy tensors for $\su(N)_0$ and $\gl(1|1)_N$ \cite{LeClair:2007aj}. Now, for a single (anti)fermion field and its (anti)ghost partner the currents $\psip\psim$, $ \betap\betam$, and $\pm\psi_{\pm}\beta_{\mp}$ together form a representation of the supercurrent algebra $\gl(1|1)_1$. A representation of $\gl(1|1)_N$ may then be obtained by simply adding N copies of these currents together - one for each Dirac fermion and ghost - so that we have representation
\begin{equation}
	\label{eqn:DGLC}
	H \equiv \psip^i\psim^i,~J \equiv \betap^i\betam^i,~\mbox{and}~S_{\pm} \equiv \pm\psi_{\pm}^i\beta_{\mp}^i~.
\end{equation}
One may check that these currents commute with the currents $L^a$, which is a necessary condition for the free stress-energy tensor to separate  into commuting parts.

The Sugawara stress-energy tensor for $\gl(1|1)_N$ is constructed from the two quadratic Casimir invariants of the supergroup \cite{Rozansky:1992np}, so that
\begin{equation}
	\label{eqn:SETSUQC}
	T_{\gl(1|1)_N} 
	 = -\frac{1}{2N}\Big(J^2 - H^2 + S_+S_- - S_-S_+\Big) 
	 + \frac{1}{2N^2}\Big(H-J\Big)^2.
\end{equation}
Applying the definitions (\ref{eqn:DGLC}) and the Wick theorem with Eqs. (\ref{eqn:FGP}), one finds the stress-energy tensor
\begin{align}
	T_{\gl(1|1)_N} 
	& = -\frac{1}{2N^2}\bigg(\psim^i\dz\psip^i + \psip^i\dz\psim^i + \betam^i\dz\betap^i - \betap^i\dz\betam^i\bigg) \notag \\
	& ~~~~~ -\frac{1}{2N^2}\bigg[(N-1)(\betap^i\betam^i)^2  - (N+1)(\psip^i\psim^i)^2  \notag \\  
	&~~~~~~~~~~ - 2(\psim^i\psip^i)(\betam^j\betap^j) + 2N(\betap^i\psim^i)(\betam^j\psip^j)\bigg]. \label{eqn:SETGL}
\end{align}
It follows immediately from Eqs. (\ref{eqn:SETCFT}), (\ref{eqn:SETSU}) and (\ref{eqn:SETGL}) that
\begin{equation}
	\label{eqn:SUS}
	T_{\textrm{free}} = T_{\gl(1|1)_N} + T_{\su(N)_0}~.
\end{equation}

Next consider $\so(N)_0$.  Let us first anticipate the form of super spin-charge separation by considering the conformal dimensions. The vector representation is N-dimensional, and the scaling dimension of a primary field is
\begin{equation}
	\Delta_{\so(N)_0} = \frac{N-1}{2(N-2)}~.
\end{equation}
In the free superconformal theory there are 4N complex field degrees of freedom, $\psi_{\pm}^i$ and $\beta_{\pm}^i$, so we expect the supercurrent algebra with which $\so(N)_0$ decomposes to have a 4-dimensional representation, 
with scaling dimension $\Delta = \Delta_{\textrm{free}}- \Delta_{\so(N)_0} = 1/[2(2-N)]$,  which should be identified with an $\osp(2|2)_N$ scaling dimension.   Now, similarly to the above, the currents $\psip\psim$, $ \betap\betam$, $\beta_\mp\beta_\mp$, $\pm\psi_{\pm}\beta_{\mp}$ and $\psi_{\mp}\beta_{\mp}$ form a representation of the supercurrent algebra $\osp$(2$|$2)$_1$, so a representation of $\osp$(2$|$2)$_N$ can be formed by the currents (\ref{eqn:DGLC}) together with
\begin{equation}
	\label{eqn:DOSPC}
	J_{\pm} \equiv \beta_\mp^i\beta_\mp^i,~\mbox{and}~\widehat{S}_{\pm} \equiv \psi^i_{\mp}\beta^i_{\mp}~.
\end{equation}
In general, the currents ($J$, $J_{\pm}$) together with $H$ generate a $\su$(2)$\otimes \uone$(1) subalgebra: in the above representation (\ref{eqn:DGLC}) and (\ref{eqn:DOSPC}), the spin $s = 1/2$ ($\beta_\pm$ are bosonic spin-1/2 fields) and the U(1) charge $b = 0$ respectively.  Notably, the representations of $\osp$(2$|$2)$_{\textrm{N}}$ are identified by $s$ and $b$ (see e.g. Refs. \cite{Maassarani:1997np, Bernard:2000vc}), with dimension $8s$ and scaling dimension
\begin{equation}
	\Delta^{b,s}_{\osp} = 2(s^2 - b^2)/(2-N)~,
\end{equation}
so it follows that the representation generated by $H$, $J$, $J_\pm$, $S_\pm$ and $\widehat{S}_\pm$ satisfies the above dimensionality and scaling weight requirements. Further, once again these currents all commute with the currents $L^a$. 

The Sugawara tensor for $\osp(2|2)_N$ is \cite{Maassarani:1997np}
\begin{equation}
\label{eqn:SETSOQC}
	 T_{\osp(2|2)_N} 
	 = \frac{1}{2(2-N)}\bigg[J^2 - H^2 - \frac{1}{2}\Big(J_+J_- + J_-J_+\Big)
	+  \Big(S_+S_- - S_-S_+\Big) + \Big(\hat{S}_-\hat{S}_+ - \hat{S}_+\hat{S}_-\Big)\bigg]~,
\end{equation}
so from the definitions (\ref{eqn:DGLC}) and (\ref{eqn:DOSPC}) and the Wick theorem, one finds
\begin{align}
	 T_{\osp(2|2)_N} 
	& = \frac{1}{2(N-2)}\bigg(\!\psim^i \dz\psip^i+ \psip^i \dz\psim^i + \betam^i\dz\betap^i - \betap^i\dz\betam^i\!\bigg) \notag\\
	& ~~~~~- \frac{1}{2(N-2)}\bigg( (\betap^i\betam^i)^2 - (\betap^i\betap^i)(\betam^j\betam^j) - (\psip^i\psim^i)^2  \notag \\ 
	&~~~~~~~~~~+ 2(\psim^i\betap^i)(\psip^j\betam^j) - 2(\psim^i\betam^i)(\psip^j\betap^j)\bigg)~. \label{eqn:SETOSP}
\end{align}
Once more, it follows immediately from Eqs. (\ref{eqn:SETCFT}), (\ref{eqn:SETSO}) and (\ref{eqn:SETOSP}) that
\begin{equation}
	\label{eqn:SOS}
	T_{\textrm{free}} = T_{\osp(2|2)_N} + T_{\so(N)_0}~,
\end{equation}
as claimed.

\section{Current-current perturbations}

In Ref. \cite{Bernard:2001np} it was shown that the left-right current-current perturbation for $\syp(2N)_0$ is marginally relevant, while the corresponding $\osp(2|2)_{-2N}$ perturbation was marginally irrelevant. The $\su(2)_0$ and $\osp(2|2)_{-2}$  case was shown in Ref. \cite{Bernard:2000vc}. Let us now verify that the left-right current-current perturbations for $\su(N)_0$ and $\so(N)_0$ are similarly marginally relevant, and those for the corresponding $\gl(1|1)_N$ and $\osp(2|2)_N$ supercurrent algebras are marginally irrelevant. We emphasise that we do not consider other possible current-current perturbations, i.e. other kinds of disorder,  beyond those corresponding to the above algebras.

Consider the SU(N) spin-charge separation (\ref{eqn:SUS}). The algebra $\gl(1|1)_N$ has two quadratic Casimirs invariants, so we add three left-right current-current perturbations to the free theory, such that the action is
\begin{equation}
	\mathcal{S} = \mathcal{S}_{\textrm{free}} + \int \frac{d^2z}{4\pi} \Big(gL^a\bar{L^a} + g_1^{\prime}(K\bar{K})_1 + g_2^{\prime}(K\bar{K})_2\Big)~.
\end{equation}
Here $L^a\bar{L^a}$ is the left-right quadratic Casimir invariant for $\su(N)_0$, with currents $L^a$ as defined in Eq. (\ref{eqn:GLAC}), and 
\begin{align}
	(K\bar{K})_1 & = J\bar{J} - H\bar{H} + S_+\bar{S}_- - S_-\bar{S}_+\notag\\
	(K\bar{K})_2 & = -(H-J)(\bar{H} - \bar{J})
\end{align}
are the two $\gl(1|1)_{N}$ left-right quadratic Casimirs. Now, for a set of operators $\{\mathcal{O}_i\}$ with couplings $g_i$, having OPEs
\begin{equation}
	\mathcal{O}_i(z,\bar{z})\mathcal{O}_j(0) \sim \frac{1}{z\bar{z}}C^k_{ij}\mathcal{O}_k~,
\end{equation}
it is well-known that the one-loop beta functions are 
\begin{equation}
	\beta_{g_k} = -\sum_{ij}C^k_{ij}g_ig_j~.
\end{equation}
As noted in Sec. (\ref{sec:PD}) the operators $L^a\bar{L^a}$ and $(K\bar{K})_{1,2}$ commute, and it follows that the beta function for $g$ decouples from those for $g_1^{\prime}$ and $g_2^{\prime}$. Applying Eq. (\ref{eqn:SUGI}), one finds that
\begin{equation} 
	\beta_g = 2Ng^2~, ~~\beta_{g_1^{\prime}} = 0~,~~\mbox{and}~\beta_{g_2^{\prime}} = -2(g_1^{\prime})^2~.
\end{equation}
Provided the couplings $g$ and $g^\prime_{1,2}$ are positive in the model under consideration, then clearly $g$ is marginally relevant, while $g_1^{\prime}$ is scale independent and $g_2^{\prime}$ is marginally irrelevant, as desired. 

For the SO(N) case (\ref{eqn:SOS}), the supercurrent algebra $\osp(2|2)_N$ has only one quadratic Casimir invariant. The perturbed action becomes 
\begin{equation}
	\mathcal{S} = \mathcal{S}_{\textrm{free}} + \int \frac{d^2z}{4\pi} \Big(gL^a\bar{L^a} + g^{\prime}(K\bar{K})\Big)~,
\end{equation}
where $L^a\bar{L^a}$ ($K\bar{K}$) is the left-right quadratic Casimir invariant for $\so(N)_0$ ($\osp(2|2)_N$), and
\begin{equation}
	(K\bar{K}) = -\Big[J\bar{J} - H\bar{H} - \frac{1}{2}(J_+\bar{J}_- + J_-\bar{J}_+)
	+  (S_+\bar{S}_- - S_-\bar{S}_+) + (\hat{S}_-\bar{\hat{S}}_+ - \hat{S}_+\bar{\hat{S}}_-)\Big]~.
\end{equation}
By identical reasoning to the SU(N) case, the beta functions for $g$ and $g^\prime$ decouple, and applying Eq. (\ref{eqn:SOGI}) one finds
\begin{equation}
	\beta_g = 2(N-2)g^2~,~~\mbox{and}~\beta_{g^{\prime}} = -4(g^{\prime})^2~,
\end{equation}
with RG behavior as expected for positive couplings. These results may also be derived (up to a normalization) from the beta functions found for a more general case \cite{Serban:2002df}, after fine-tuning the parameters such that the perturbations have the correct symmetries.

\section{Density of states exponents}
\label{sec:DSE}

A useful example is to consider the perturbation
\begin{equation}
	\mathcal{S}_E = \int \frac{d^2z}{4\pi} ~ E \,  \rho(z,\bar{z})
\end{equation}
where $E$ is an energy and $\rho(z,\bar{z})$ is the density of states
\beq
	\rho(z,\bar{z})
	 \equiv \bar{\psi}_-^i(\bar{z})\ps{+}{i}{z} + \ps{-}{i}{z}\bar{\psi}_+^i(\bar{z}) + \bar{\beta}_-^i(\bar{z})\be{+}{i}{z} 
	+ \be{-}{i}{z}\bar{\beta}_+^i(\bar{z})~. \label{eqn:DDS}
\eeq
Let $\rho$ have scaling weight $\Gamma$ near a fixed point. Then since the action must be conformally invariant, the energy $E$ must have scaling weight $2 -\Gamma$ and it follows that the density of states exponent, $\nu$, is
\begin{equation}
	\langle\rho(E)\rangle \sim E^{\nu},~ \nu \equiv \Gamma/(2-\Gamma)~.
\end{equation}

The scaling dimension $\Gamma = 2 \Delta$,  where $\Delta$ is the conformal dimension for the appropriate $4$-dimensional representation of $\gl(1|1)_N$ or $\osp(2|2)_{N,-2N}$, which are known.   
We can give an independent, self-contained derivation of $\Gamma$ as follows. Suppose there is a spin-charge separation $T_\textrm{free} = T_\mathfrak{g'} + T_\mathfrak{g}$, where $T_{\mathfrak{g}}$ is a stress-energy tensor for the massive degrees of freedom $\su(N)_0, \syp(2N)_0$ or $\so(N)_0$, and $T_\mathfrak{g'}$ is the stress energy tensor for the supercurrent algebra, with
\begin{equation}
	\label{eqn:GSCA}
	T_\mathfrak{g'} = -\kappa \bigg(\psim^i\dz\psip^i + \psip^i\dz\psim^i + \betam^i\dz\betap^i - \betap^i\dz\betam^i\bigg) + \ldots
\end{equation}
where the quartic terms are contained in the ellipsis. From arguments given above, the scaling dimension  of $\rho$ near a fixed point may be found by considering only the OPEs $T_\mathfrak{g'}(z)\rho(w,\bar{w})$ and $\bar{T}_\mathfrak{g'}(\bar{z})\rho(w,\bar{w})$: by definition the (anti)holomorphic conformal weight and hence scaling dimension may be calculated by finding the $(z-w)^{-2}$ ($(\bar{z}-\bar{w})^{-2}$) terms in these OPEs.  From Eq. (\ref{eqn:DDS}) it can clearly be seen that only the kinetic terms of $T_\mathfrak{g'}$, shown in Eq. (\ref{eqn:GSCA}), can produce such terms, so that one finds
\begin{equation}
	 T_\mathfrak{g'}(z)\rho(w,\bar{w}) \sim \frac{\kappa}{(z-w)^2}\rho(w,\bar{w}) + \ldots
\end{equation}
and similarly for the antiholomorphic case. Hence $\Gamma = 2\kappa$.

For the cases $\mathfrak{g} = \syp(2N)$, $\su(N)$ and $\so(N)$ we may now simply calculate $\Gamma$ and $\nu$, by extracting the appropriate $\kappa$ coefficient from the stress-energy tensors written explicitly in Eqs. (\ref{eqn:SETGL}) and (\ref{eqn:SETOSP}). Note that for the $\syp(2N)$ case (\ref{eqn:SPS}), the coefficient $\kappa$ is found from Eq. (\ref{eqn:SETOSP}) by substituting $N \to -2N$. Results are shown in Table \ref{tab:DS}. Note that the $\su(2)$ value $\nu =1/7$ agrees with previous results \cite{Bernard:2000vc}.  Interestingly, for the SO(N) case, the density of states diverges for $N\geq 2$.  

\begin{table}[ht]
	\begin{tabular*}{0.45\textwidth}{@{\extracolsep{\fill}}ccc}
		\hline
		\hline
 		$\mathfrak{g}$ & $\Gamma$ & $\nu$ \\
 		\hline
 		$\syp(2N)$ & $1/[2(N+1)]$ & $1/(4N +3)$\\
		$\su(N)$ & $1/N^2$ & $1/(2N^2 -1)$ \\
		$\so(N)$ & $1/(2-N)$ & $1/(3-2N)$ \\
		\hline
		\hline
	\end{tabular*}
	\caption{Density of states scaling dimensions 
 and associated exponents for different symmetry groups.}
	\label{tab:DS}
\end{table}

\bibliography{spinchargesep}

\end{document}